\documentclass[aip,amsmath,amssymb,graphicx,preprint]{revtex4-1}

\usepackage{graphicx}
\usepackage{dcolumn}
\usepackage{bm}

\usepackage[utf8]{inputenc}
\usepackage[T1]{fontenc}
\usepackage{mathptmx}
\usepackage{etoolbox}

\draft 

\makeatletter
\def\@email#1#2{%
 \endgroup
 \patchcmd{\titleblock@produce}
  {\frontmatter@RRAPformat}
  {\frontmatter@RRAPformat{\produce@RRAP{*#1\href{mailto:#2}{#2}}}\frontmatter@RRAPformat}
  {}{}
}%
\makeatother

\begin{document}

\preprint{}

\title{All field emission models are wrong, ... but are any of them useful ?} 



\author{Anthony Ayari}
\author{Pascal Vincent}
\author{Sorin Perisanu}
\author{Philippe Poncharal} 
\author{Stephen T. Purcell}
\email[]{anthony.ayari@univ-lyon1.fr.}
\affiliation{Univ Lyon, Univ Claude Bernard Lyon 1, CNRS, Institut Lumi\`ere Mati\`ere, F-69622, VILLEURBANNE, France.}


\date{\today}

\begin{abstract}
Field emission data are often represented on a Fowler-Nordheim plot, but a new empirical equation has been recently proposed to better analyse experiments. Such an equation is based on approximations of the Murphy and Good model and predicts that a constant parameter $\kappa$, depending only on the work function of the emitter, can be extracted from the data. We compared this empirical equation with simulations of the Murphy and Good model in order to determine the range of validity of the approximations and the robustness of the relationship between $\kappa$ and the work function. We found that $\kappa$ is constant only over a limited range of electric fields and so depends significantly on the field enhancement factor. This result calls into question the usefulness of the new empirical equation.

\end{abstract}


\maketitle 

\section{Introduction}

The Fowler-Nordheim (FN) \cite{fowler1928electron,jensen2017introduction} plot is a graphical representation of IV data widely used even beyond the vacuum electron source domain, for example in molecular electronics\cite{beebe2006transition}, superconducting devices\cite{super} and semiconducting diodes\cite{di2012metal}, to cite a few. In principle, and often in practice, the plot is linear from which two experimental coefficients (the slope and intercept) can be extracted. The fact that the FN plot gives "close to" a straight line is probably why it is so generally used, but other representations of the data  also give "close to" a straight line. We used a vague term as "close to" in quotation marks because a widespread problem among many experimental reports is that a linear fit is forced on IV data showing significant deviations from a straight line. A departure from the linear behavior is a clear indication that additional physical mechanisms come into play, for examples a fuller development of the theory or even a voltage drop coming from a series resistance. Theoretically, the FN plot should be strictly linear only in the case of a triangular tunneling barrier. However, the presence of an exponential in the expression of the current makes it very difficult to observe the curvature induced by a more physically correct barrier including an image charge. Another important issue is that the FN model needs 3 independent physical quantities, but the fit gives only 2 parameters. So, the usefulness of the FN model is questionable, if meaningful effective physical quantity are sought. 

Recently, a new analytical form of the field emission current, and a method to test it, have been proposed\cite{forbes2008call}. The method has begun to attract some interest from the field emission community\cite{popov2018experimental,zubair2018fractional,forbes2021pre}. However, the reliability of the information that can be deduced from this model has not really been studied very thoroughly, in particular the value of the exponent $\kappa$. “All models are wrong, but some are useful” is a famous quote often attributed to the British statistician George E. P. Box. The main goal of this article is to test if this new analytical formula is useful in field emission. In particular, we present some numerical calculations based on the more general Murphy and Good model\cite{murphy1956thermionic} in order to determine the validity of the approximations and the uncertainty it induces in the determination of the work function. In section \ref{difmod}, the different models and numerical methods that will be used are presented. In section \ref{sectiondata}, we show that different models predict different exponent $\kappa$ for a given work function and we propose a new relationship between $\kappa$ and the work function. However, it will be demonstrated that such relationship is useful in a limited range of electric field because $\kappa$ is not constant on the full range of electric field for field emission. Finally, some general comments are made in section \ref{secfin} to explain why a method based only on the derivative of the current leads to some difficulty in estimating physical parameters in field emission. A new method based on the measurement of the current, its first and second derivative is then proposed.

\section{which model to choose?}\label{difmod}
In most transport measurements, the current I flowing through a device is measured as a function of the applied voltage V. In the case where the transport mechanism is dominated by a tunnel barrier, and that the barrier shape changes with the voltage, an exponential like dependence of I as a function of -1/V is often observed. However the expression of the current might be far from a pure exponential. In field emission, several mathematical expressions of the field emission current can be found for instance in ref. \onlinecite{jensen2017introduction,modinos1984secondary} or in table 1 of ref. \onlinecite{forbesa2015jordan}. It is then crucial to decide which theory to compare with experiments. For an experimentalist, the first answer to this question might be to chose the simplest theory that fits the data. Unfortunately, this choice of the FN theory with a triangular barrier doesn't give meaningful information about the emitter as it strongly underestimates the value of the current density. Another strategy is to increase step by step the complexity level of the theory and discard the one that cannot permit to extract reliable physical parameters from the experiments. 

The choice in the theory raises another question. What is a reliable physical parameter ? Three physical quantities can be defined : the work function $\phi$, the emission area S and the field enhancement factor $\beta$ relating the applied voltage to the electric field.  However these physical quantities are most often not uniform over the emitter, not constant in the measurement time or the applied voltage, and not independent of each other. As a minimum average values must be considered. In the highest level of complexity, the theory needs to take into account the exact geometry of the emitter\cite{de2021using} and its atomic structure\cite{lepetit2019three}. It is also very demanding on the experimental side because it is often difficult to know if the electron emission comes from the whole apex surface or only from a nanoprotusion with a much lower radius of curvature.
\subsection{The Murphy and Good theory}
Murphy and Good\cite{murphy1956thermionic} proposed a model based on several reasonable hypothesis where $\phi$, S and $\beta$ are constants. The current can be expressed as an integral on easily numerically calculable functions and is given by Eq. 19 in Ref.\onlinecite{murphy1956thermionic} : 
\begin{equation}\label{eqfufllMG}
I = \frac{S}{2\pi^2}\frac{mek_BT}{\hbar^3}\left[\int_0^{W_l} D(W,F)\ln\left[1+\exp\left(\frac{\mu-W}{k_BT}\right)\right]dW+\int_{W_l}^\infty \ln\left[1+\exp\left(\frac{\mu-W}{k_BT}\right)\right]dW\right]
\end{equation}
where $m$ is the electron mass, $e$ is the electron charge, $k_B$ the Boltzmann constant, $T$ the temperature, $\hbar$ is the Planck constant,   $W$ is the electron energy incident on the barrier, $F$ is the electric field, $\mu$ the Fermi energy,
\begin{equation}
W_l = \phi + \mu -\frac{1}{\sqrt{2}}\frac{e^3F}{4\pi\varepsilon_0}
\end{equation}
where $\varepsilon_0$ is the vacuum permittivity and the transmission is :
\begin{equation}
D(W,F) = \frac{1}{1+exp(Q(W,F))}
\end{equation}
where
\begin{equation}
Q(W,F) = b\frac{(\phi + \mu -W)^{3/2}}{F}v(y)
\end{equation}
where v(y) is the barrier shape correction factor that depends on the applied electric field through the variable y and can be expressed as a combination of elliptic integrals.

The model takes into account the effects of the temperature and the image charge potential. It provides an expression of the current after a Taylor expansion (see Eq. 56 in ref. \onlinecite{murphy1956thermionic} or eq. 1.54 in ref. \onlinecite{modinos1984secondary}) that can be further simplified, in order to give a more compact expression by discarding the temperature dependence and an almost constant preexponential term usually expressed as t(y):
\begin{equation}\label{eqFNMG}
    I = aS\frac{(\beta V)^2}{\phi}\exp(-bv(y)\frac{\phi^{3/2}}{\beta V})
\end{equation}

where a and b are terms that depend only on universal physical constants. This simplified model has two main drawbacks: the formula is still rather complicated and the expression of v(y) is even more complicated. Fortunately (or unfortunately), in the case of a triangular barrier v(y) = 1 and then the equation predicts that plotting the current in Fowler-Nordheim coordinates, (i.e. representing the logarithm of $I/V^2$ as a function of $1/V$) gives a straight line. It is unfortunate because, plotting some data described by Eq. \ref{eqFNMG} in Fowler-Nordheim coordinates gives a plot that looks like a straight line too. It is then very tempting to forget about the v(y) term whereas neglecting v(y) in the exponential leads to an overestimation of S by several orders of magnitude.  

\subsection{\label{eq Forbes}A profusion of analytical expressions}
New analytical formulae have been proposed in the past few years in order to better describe experimental data for a potential barrier with an image charge \cite{forbes2008call}, for small radius emitters\cite{kyritsakis2015derivation}, for temperature dependent emission\cite{jensen2019reformulated}, for rough surfaces\cite{zubair2018fractional} or for specific materials such as nanotubes\cite{liang2008generalized} or graphene\cite{forbes2010thin}. These formulae can essentially be distinguished by the coefficient of the power law they predict. 

Here, we will be essentially interested by the analytical approximation of v(y) used in Ref. \onlinecite{forbes2008call} as it gives a more accurate simplification of Eq. \ref{eqFNMG} than the triangular barrier approximation. It was obtained for the case of a tunneling barrier with a classical image charge correction and can be expressed as (Eq. 7 in Ref. \onlinecite{forbes2008call}):
\begin{equation}\label{eqForb}
    I = a_\kappa S\frac{(\beta V)^\kappa}{\phi}\exp(-b\frac{\phi^{3/2}}{\beta V})
\end{equation}
where $\kappa$ and $a_\kappa$ are independent of V and $\beta$ but vary with the work function. 
\begin{equation}\label{ak}
    a_\kappa = \frac{a}{\phi}\exp\left(b\frac{\phi^{3/2}}{F_\phi}\right)F_\phi^{\eta/6}
\end{equation}
\begin{equation}\label{Fp}
   F_{\phi} = \frac{4\pi\varepsilon_0\phi^2}{e^3}
\end{equation}
\begin{equation}\label{eta}
\eta = \frac{b\phi^{3/2}}{F_{\phi}}
\end{equation}
\begin{equation}\label{kappa}
\kappa = 2-\eta/6
\end{equation}

Now, a plot of the logarithm of $I/V^\kappa$ as a function of 1/V (where $\kappa \sim $1.2 for a tungsten field emitter) is really a straight line and should give a better fit of the experimental data. Unfortunately, as in general $\kappa$ is unknown, such an exact plot cannot be directly obtained.  It was then proposed either to plot the logarithm of $I/V^k$ as a function of 1/V with different $k$ values to check which one gives the best fit to the data (the k value corresponding to the best fit should then be equal to $\kappa$) or to obtain the voltage derivative of the current, because the new analytical approximation gives the following expression:

\begin{equation}\label{theformule}
    \frac{V^2}{I}\frac{dI}{dV} = b\frac{\phi^{3/2}}{\beta}+ \kappa V
\end{equation}

So a plot of this ratio as a function of V should be linear with a slope giving directly the value of $\kappa$. Field emission is a strongly non-linear phenomenon and so the impact of the various approximations in the Murphy and Good model on the robustness of this linear equation needs to be tested.

\subsection{Numerical methods of calculations}\label{nummet}

Two appealing aspects of Eq. \ref{eqForb} are that it offers a new way of analysing field emission data and it gives a simple method to extract the work function independently of the value of $\beta$. However, as the proposed analytical formulae in Ref. \onlinecite{forbes2008call} are based on some approximations, it is first necessary to estimate the level of uncertainties it induces. In ref. \onlinecite{forbes2007reformulation}, it was shown that the analytical approximation of v(y) has a relative error better than $0.33\%$. Such an error is very good, however as v(y) is in an exponential, the error in the current is much higher. In Ref. \onlinecite{forbes2008call}, it was proven that a linear relationship such as in Eq. \ref{theformule} was also valid for a model with the analytical Murphy and Good formula. It was also noted a $8.3\%$ discrepancy  on the value of $\kappa$ giving 1.1 instead of 1.2 for a work function of 4.5 eV. Such a discrepancy was further studied for various models\cite{forbes2021pre} but its impact on the accuracy of the extraction of physical parameters was not quantified. For instance, from an experimentalist point of view, it is not clear how accurate should be the measure of $\kappa$ to have a good estimate of the work function and what is the uncertainty on $\phi$ due to the $8.3\%$ discrepancy in $\kappa$.

We performed numerical  simulations using NumPy and SciPy Python packages where elliptic functions required for the calculation of v(y) are implemented. The data and programs used in this article can be found in the Zenodo data repositiory\cite{ayarizeno}. As an illustrative example, the calculation were done for a single field emitter with an emission area of 604 $nm^2$, $\beta = 0.0015 nm^{-1}$ and the voltage was varied between 1300 and 1600 V with 10 V increments. These voltages correspond to a range of low fields in order to allow some comparisons with some preliminary experimental data at low current presented at the IVNC2021 conference\cite{ayari2021does}. These data will be further analysed in an upcoming article. At the end of the article, the value of the voltage will be extended to cover the whole range of  typical electric fields. Several values of the work function will be tested. Three levels of complexity in the models were selected. 

In the first level in decreasing order of complexity, the full Murphy and Good formula as given by Eq. \ref{eqfufllMG} was used. Beside the WKB approximation, it is based on exact calculations. The integration were performed numerically by the rectangle rule with an energy excursion range of 3 eV around the Fermi energy and a step of integration of 10 meV (an energy excursion range of 4 eV and a 10 $\mu$eV step were also tested with no significant change in the results). The calculations were performed for 0 K and 300 K. The derivative of the current was calculated analytically without approximation and then directly numerically integrated from the obtained formula in order to avoid the uncertainty encountered in Ref. \onlinecite{forbes2021pre} by their local gradient method :
\begin{equation}
\frac{\partial I}{\partial V} = \frac{S}{2\pi^2}\frac{mek_BT}{\hbar^3}\left[\int_0^{W_l} \ln\left[1+\exp\left(\frac{\mu-W}{k_BT}\right)\right]\frac{\partial D(W,F)}{\partial V}dW\right]
\end{equation}

with 

\begin{equation}
\frac{\partial D(W,F)}{\partial V} = D(W,F)^2exp(Q(W,F))\frac{\partial Q(W,F)}{\partial V}
\end{equation}
and 
\begin{equation}
\frac{\partial Q(W,F)}{\partial V} = \beta\frac{\partial Q(W,F)}{\partial F}=\beta b\frac{(\phi + \mu -E_{//})^{3/2}}{F}s(y)
\end{equation}
where
\begin{equation}
s(y) = v(y)-\frac{y}{2}\frac{\partial v(y)}{\partial y}
\end{equation}

The only equation that cannot be found in Ref.\onlinecite{murphy1956thermionic} is the expression of $\partial v(y)/\partial y$ :
\begin{equation}
\frac{\partial v(y)}{\partial y} = -\frac{3}{2}\frac{y}{\sqrt{1+y}}\ K\left(\frac{\sqrt{1-y}}{\sqrt{1+y}}\right)
\end{equation}
for y < 1 and 
\begin{equation}
\frac{\partial v(y)}{\partial y} = -\frac{3}{2}\frac{\sqrt{y}}{\sqrt{2}}\ K\left(\frac{\sqrt{y-1}}{\sqrt{2y}}\right)
\end{equation}
for y > 1, where K is the complete elliptic integral of the first kind expressed as a function of its elliptic modulus. 

In the intermediate level of complexity, Eq. \ref{eqFNMG} was used. In the third level of complexity, we calculated Eq. \ref{eqForb}. Comparing the second and third level of complexity allows to identify the impact of the approximation of v(y) alone. Comparing the first and third level of complexity allows to determine the difference between the full model including temperature effects and its simplified version. We will not consider the case where $\beta$, S or $\phi$ depend on V as in Ref. \onlinecite{forbes2021pre} because without these complications, the theoretical situation requires already some clarifications and also because experimentally, there is still some hope that for certain experiments a constant value of $\beta$, S and $\phi$ is possible, for example, by measuring the central current of a sufficiently large facet through a probe hole for a metallic emitter.
\section{Relationship between $\phi$ and $\kappa$}\label{sectiondata}

\subsection{The value of $\kappa$ at 4.6 eV}

In this part, a work function of 4.6 eV is chosen in order to allow a comparison with the calculations in Ref. \onlinecite{forbes2021pre}.  As we will see, their results seem slightly inaccurate. They used a software more adapted for experiment control than for data analysis or numerical simulations, so the algorithm of optimisation might be less precise. We will first test the consistency of our calculations and analysis with the simplest model because there exists an analytical expression for $\kappa$ that can be compared with the value extracted from the fits. The analytical relationship between the work function and $\kappa$ is given in the simplest model by combining Eq. \ref{kappa}, \ref{eta} and \ref{Fp}:
\begin{equation}\label{giveK}
\kappa = 2-\frac{e^3b}{6\times4\pi\varepsilon_0\sqrt{\phi}}
\end{equation}

Mathematically $\kappa$ can vary between 0 and 2 and physically a value between 0.8 and 1.5 largely covers the range of reasonable work function values. According to this analytical formula we have $\kappa =$ 1.2356383319707134 for $\phi=$ 4.6 eV. Such a level of accuracy is not necessary in practice and will be reduced later, but is just presented here for comparison with the fitting method. 

As explained in the paragraph \ref{eq Forbes}, a first method to extract $\kappa$ is by multiple fitting. To avoid any confusion, we will use the notation $\kappa_m$ when $\kappa$ is extracted by this method. We calculated the current using eq. \ref{eqForb} (with a work function of 4.6 eV and the parameter values given in \ref{nummet}) and performed linear fits of the logarithm of $I/V^k$ as a function of 1/V for different values of $k$. The minimum  value of the least squares linear fits is obtained for $\kappa_m =$1.2356383. So contrary to Ref. \onlinecite{forbes2021pre}, we see that this method can determine $\kappa$ with an accuracy higher than 6 significant digits. Another method is by fitting the ratio in Eq. \ref{theformule}. We will use the notation $\kappa_r$ when $\kappa$ is extracted by this fit. We obtained $\kappa_r =$1.235638331970699 $\pm1\times10^{-14}$. The local gradient method in Ref. \onlinecite{forbes2021pre} gave 1.28. The local gradient method is very demanding in terms of voltage step size, whereas with our method an excellent accuracy can be obtained although our voltage steps are large (i.e. 10 V). So numerically, at this stage, there is no reason to prefer the multifit method compared to the method based on the ratio in Eq. \ref{theformule}.

\begin{figure}
\includegraphics{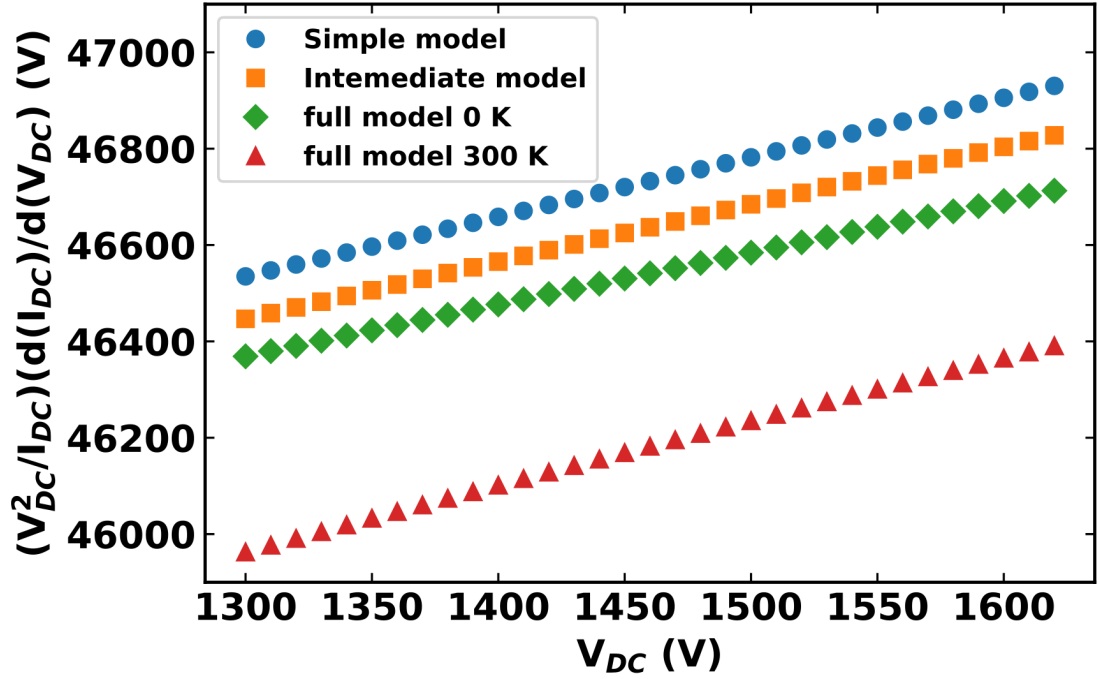}%
\caption{Variation of the expression of the left hand side of equation \ref{theformule} as a function of the applied DC voltage. The circles correspond to the simplified model from ref. \onlinecite{forbes2008call}, the squares correspond to the analytical formula in Murphy and Good (Eq. \ref{eqFNMG}), the diamonds correspond to the full Murphy and Good model at 0 K and the triangles correspond to the full Murphy and Good model at 300 K. \label{ratios}}%
\end{figure}

The current calculated with the intermediate level of complexity gave a value of $\kappa_m =\kappa_r =$1.19 for both methods. This value is equivalent to a work function of 4.1 eV according to Eq.\ref{giveK}. The uncertainty in the fit of the ratio is higher as $\kappa_r$ now is known with an imprecision of $\pm1.7\times10^{-4}$, meaning that the curve has a slight non-linearity. Thirdly, when the current is calculated with the more complex model at 300 K, the multifit method gives $\kappa_m =$1.338 and the ratio method gives $\kappa_r =$1.335$\pm0.003$. This value is equivalent to a work function of 6.1 eV according to Eq.\ref{giveK}. The results are summarized in table \ref{tab:table1}. It seems clear that using Eq.\ref{giveK} is not reliable to extract the work function of the emitter and it was probably not intended for that purpose. However what is remarkable is that the linearity in voltage in Eq.\ref{theformule} is sufficiently valid even with the more complex model as shown in Fig. \ref{ratios} and the values calculated by the different models are within 1$\%$. It can be also seen that the value of the constant term in Eq.  \ref{theformule} or for the other models is much bigger than the expected variations in this voltage range. Extracting a reliable value of the slope might be difficult experimentally in the presence of noise.

\begin{table}
\caption{\label{tab:table1} This table gathers the different values of the exponent $\kappa$ in the prefactor of the current and the apparent work functions obtained for the three different models and three different extraction methods for a fixed work function of 4.6 eV.}
\begin{ruledtabular}
\begin{tabular}{l|c|c|c}
Model& $\kappa_m$\footnote{obtained by the multifit method for different values of k.}&$\kappa_r$\footnote{obtained by extracting the slope from a fit with Eq. \ref{theformule}.}& \text{Work function (eV)}\footnote{obtained from eq. \ref{giveK} and the value of $\kappa_r$ of the corresponding model.}\\
\hline
Full model 300 K\footnote{The current was calculated with eq.\ref{eqfufllMG}} & 1.338 & 1.335$\pm0.003$&6.1\\
\hline
Intermediate model\footnote{The current was calculated with eq.\ref{eqFNMG}} & 1.19 & 1.1902 $\pm1.7\times10^{-4}$&4.1\\\hline
Simple model\footnote{The current was calculated with eq.\ref{eqForb}} & 1.2356383 & 1.235638331970699 $\pm1\times10^{-14}$&4.6\\
\end{tabular}
\end{ruledtabular}
\end{table}

\subsection{The value of $\kappa$ in the low field range}

If the Murphy and Good model is correct but the simple analytical equation does not allow to extract a reliable value of the work function, it is then necessary to give a new relationship between $\phi$ and $\kappa$. The simulations with the full Murphy and Good model but for a work function of 4.62 eV gives a fitted ratio with  $\kappa_r =$1.339$\pm0.003$. So, theoretically, the uncertainty in the fit corresponds to an uncertainty in the work function below 20 meV. It can be concluded that an experimental measurement of $\kappa$ with an accuracy of 2 significant digits after the decimal point is enough to determine the work function with an accuracy better than 0.1 eV. This corresponds to an uncertainty of about 2$\%$ in the determination of the work function which we consider as good enough for practical applications. 

The new relationship between $\phi$ and $\kappa$ can be obtained by varying the work function of the emitter. Nevertheless, it is necessary to be careful about the voltage range where the calculations are done. For a tungsten emitter, it is sometimes considered that the typical range of electric field is between 3 and 7 V/nm (see for instance p.98 in ref. \onlinecite{dyke1956field}). The low part of this range depends on the practical minimal current that can be detected in an experiment. The high part is determined by the maximum current an emitter can sustain before heating effects take place. It is strongly related to the transparency of the barrier and so to the value of the work function. A low workfunction emitter needs less electric field to have its tunneling barrier fully transparent. The range of electric field can then be renormalized by the effect of the work function\cite{forbes2013development}. In this part, we kept the range of the ratio between the electric field and $F_\phi$ between 0.133 and 0.165. These values are rather low but as explained above it is due to the fact that our experiments in ref. \onlinecite{ayari2021does} where performed at low current (i.e. below the pA range). However, if such an experiment were to be truly performed, it would be the minimum and maximum currents that would be kept constant and not the renormalized field\cite{popov2019electrical}. So, we performed also the same calculations with fixed minimum and maximum currents.

\begin{figure}
\includegraphics{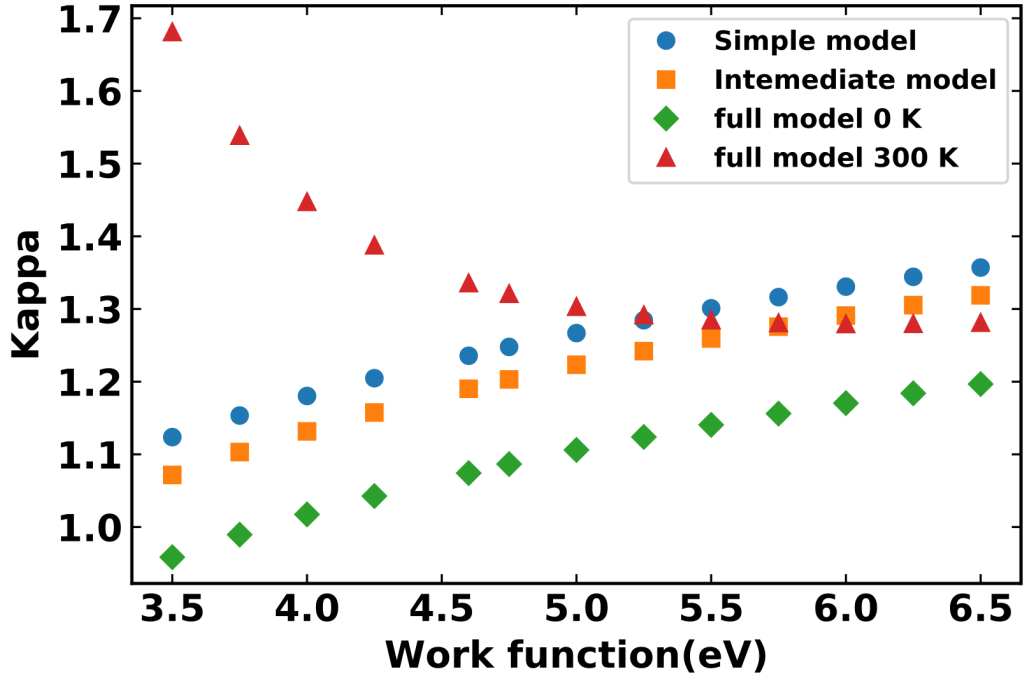}%
\caption{Variation of the value of $\kappa_r$ as a function of the work function. The circles correspond to the simplified model from ref. \onlinecite{forbes2008call}, the squares correspond to the analytical formula in Murphy and Good (Eq. \ref{eqFNMG}), the diamonds correspond to the full Murphy and Good model at 0 K and the triangles correspond to the full Murphy and Good model at 300 K. \label{KP}}%
\end{figure}

For the case where $\phi$ is varied between 3.5 eV and 6.5 eV with the same range of reduced field. We observed that the different models have very different predictions (Fig.\ref{KP}). For a given value of $\kappa$, the simple model underestimates the work function by roughly 0.5 eV compared to the intermediate model and by 1.75 eV compared to the full model at 0 K. However these 2 models have a similar behavior and can both be fitted by an expression similar to eq. \ref{giveK} :

\begin{equation}\label{giveK2}
\kappa = A+\frac{B}{\sqrt{\phi}}
\end{equation}

where for the full model at 0 K, $A \approx 1.85$ (instead of 2 for eq. \ref{giveK}) and $B \approx -1.67$ (instead of -1.64 for eq. \ref{giveK}). Although the coefficients of the more complex models are relatively close to the coefficients of the simple model, the impact on the predicted work function is significant (see table \ref{tab:table2}).

The full model at room temperature has a drastically different behavior. $\kappa_r$ decreases as a function of the work function between 3.5 eV and 5.5 eV instead of increasing for the other models (Fig. \ref{KP}) and it is roughly constant in the higher range of work function. It is then much harder at room temperature to distinguish high values of the work function. The range of possible value of $\kappa$ is very narrow and most probably it will be around 1.3 whatever the sample. So the experiments should preferentially be performed at low temperature if the emitter has a middle to high work function in this range of field. For low work function materials, measurements at 300 K seem to have a better capability to separate different work functions. In any case, Eq. \ref{giveK} should not be used to determine the work function.

For the case where $\phi$ is varied between 3.5 eV and 6.5 eV with the same range of current, the results are rather similar and we just present in table \ref{tab:table2} the coefficients of the fits.

\begin{table}
\caption{\label{tab:table2} Coefficients of the fit of $\kappa_r$ with Eq. \ref{giveK2} for the three different models and with fixed normalized field (A,B) or fixed current range (A$_I$, B$_I$).}
\begin{ruledtabular}
\begin{tabular}{l|c|c|c|c}
Model& A & B & A$_I$ & B$_I$\\
\hline
Full model 0 K\footnote{The current was calculated with eq.\ref{eqfufllMG}} & 1.85 & -1.67&1.89 & -1.76\\
\hline
Intermediate model\footnote{The current was calculated with eq.\ref{eqFNMG}} & 2 & -1.74& 1.97 & -1.67\\\hline
Simple model\footnote{The current was calculated with eq.\ref{eqForb}} & 2 & -1.64 &2 & -1.64\\
\end{tabular}
\end{ruledtabular}
\end{table}

\subsection{The value of $\kappa$ in the full range of field}

The results in the preceding paragraph showed that approximations of the Murphy and Good theory have a significant impact on the value of $\kappa$ when the work function is varied. It is then important to check the influence of the electric field on $\kappa$. After all the simple model predicts that $\kappa$ should be constant but this might not be true for the full Murphy and Good theory. In fact, a slight non-linearity was already suspected in ref.\onlinecite{forbes2021pre}.

\begin{figure}
\includegraphics{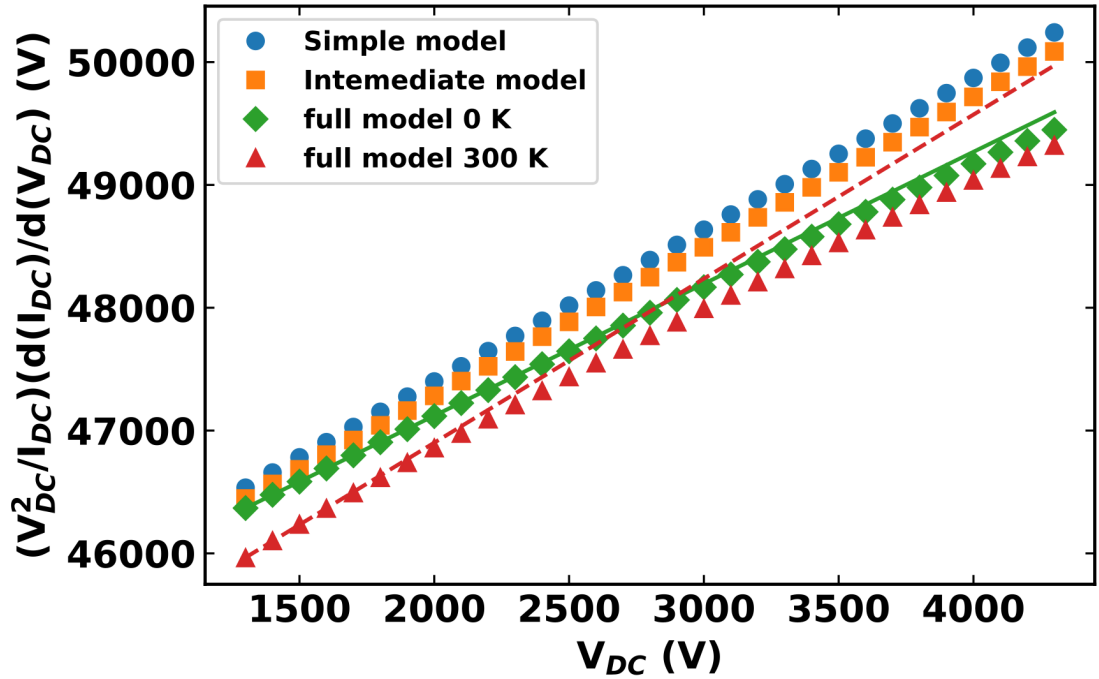}%
\caption{Variation of the value of $\kappa_r$ as a function of the applied DC voltage for a large range. The circles correspond to the simplified model from ref. \onlinecite{forbes2008call}, the squares correspond to the analytical formula in Murphy and Good (Eq. \ref{eqFNMG}), the diamonds correspond to the full Murphy and Good model at 0 K and the triangles correspond to the full Murphy and Good model at 300 K. The solid line is a linear fit of the full Murphy and Good model at 0 K at low voltage. The dashed line is a linear fit of the full Murphy and Good model at 300 K at low voltage.\label{KPHF}}%
\end{figure}

The calculations were performed for $\phi =$ 4.6 eV and a voltage between 1300 and 4330 V corresponding to a range of reduced field between 0.13 and 0.43 (i.e. 1.95 V/nm to 6.45 V/nm) that covers roughly the whole experimentally accessible range of fields. As expected theoretically, the simple model shows on Fig.  \ref{KPHF} a linear behavior of the ratio in Eq.\ref{theformule} on the full range of voltage. The intermediate model is almost linear. The value of the slope $\kappa_r$ goes from 1.19 at low field to 1.23 for high field. Nevertheless, even such a small change has important consequences on the predicted value of the work function. As in the low field range a value of 1.23 corresponds to a value of 5.1 eV instead of 4.6 eV. A non-linear deviation is visible for the full model at 0 K and is particularly important at 300 K. The extracted value of $\kappa_r$ in the high field range may differ by more than 10$\%$ which has a strong impact on the deduced value of the work function.

At 0 K, for the full Murphy and Good model, the value of $\kappa_r$ goes from 1.07 at low field to 0.92 for high field. Such a value would correspond to a work function below 3.5 eV at low field. At 300 K, for the full Murphy and Good model, the value of $\kappa_r$ goes from 1.33 at low field to 0.94 for high field. Such a low value is simply impossible at low field. As the full Murphy and Good model is supposed to be more accurate than the simple model, it can be deduced from this calculations that $\kappa$ cannot be considered as constant in the typical range of electric fields in experiments. This means that if this type of measurements are performed, a knowledge of $\beta$ is necessary to extract the value of $\phi$. This seems to limit the interest of this method that was precisely proposed in order to separate the extraction of $\phi$ and $\beta$ from experimental data.

\section{Is it possible to separate $\phi$ and $\beta$?}\label{secfin}

\subsection{General remarks}
\begin{figure}
\includegraphics{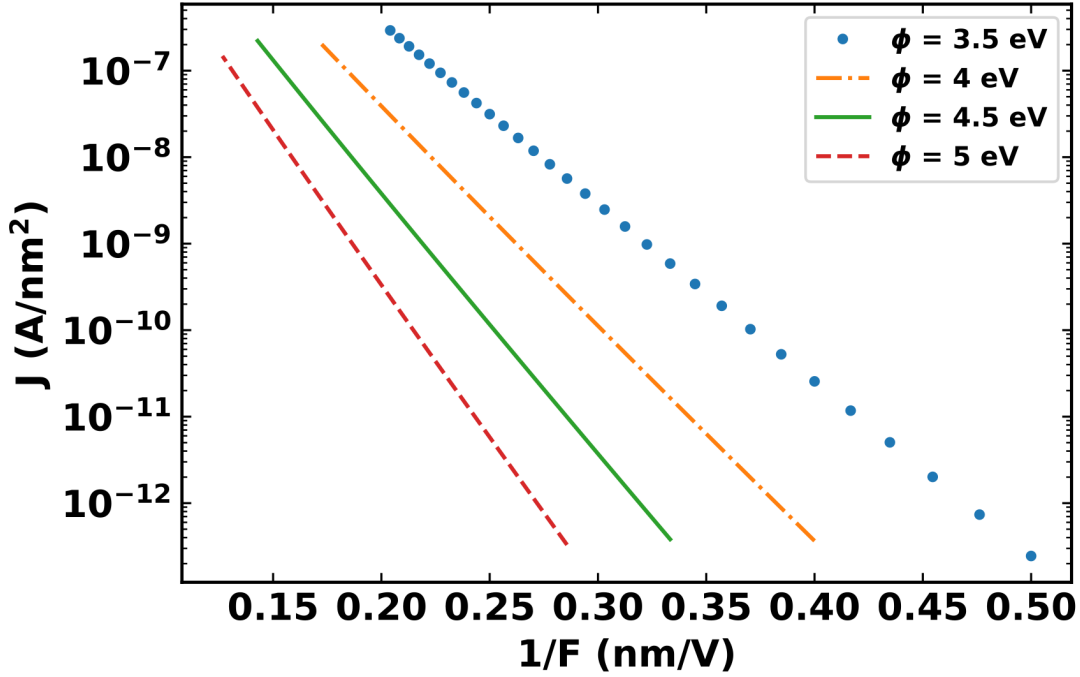}%
\caption{Field emission current density calculated from eq. \ref{eqfufllMG} at 300 K for 4 different work functions.\label{figurereferee1}}%
\end{figure}
\begin{figure}
\includegraphics{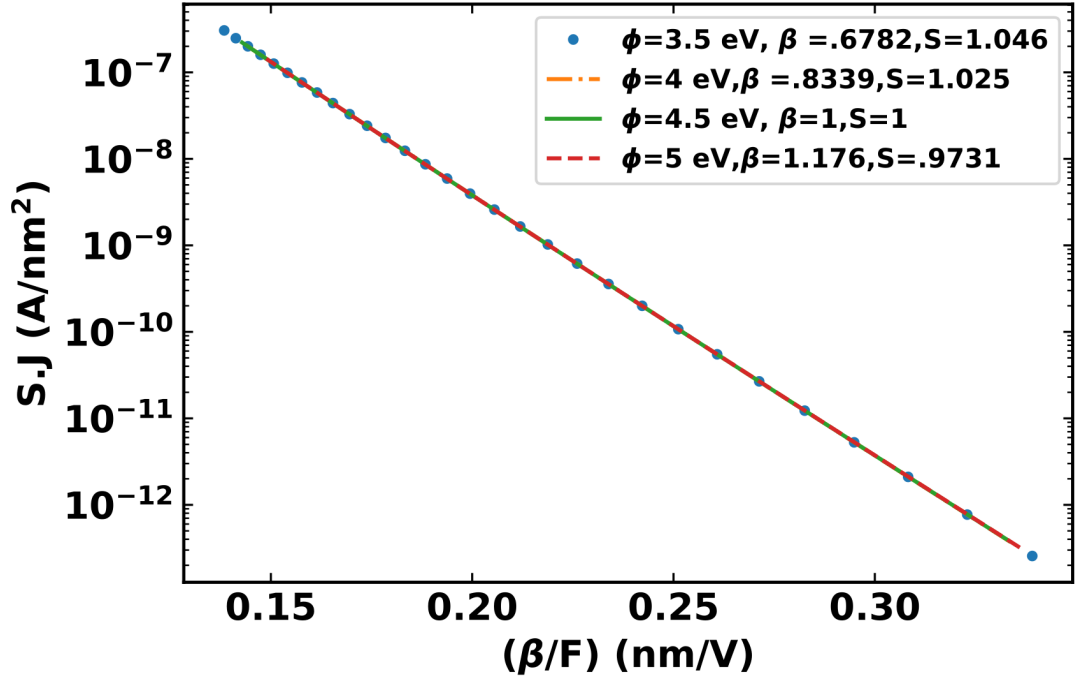}%
\caption{Field emission current as a function of the inverse of the voltage calculated from eq. \ref{eqfufllMG} at 300 K for 4 different work functions, with S and $\beta$ chosen to fit the curve at 4.5 eV. \label{figurereferee2}}%
\end{figure}
\begin{figure}
\includegraphics{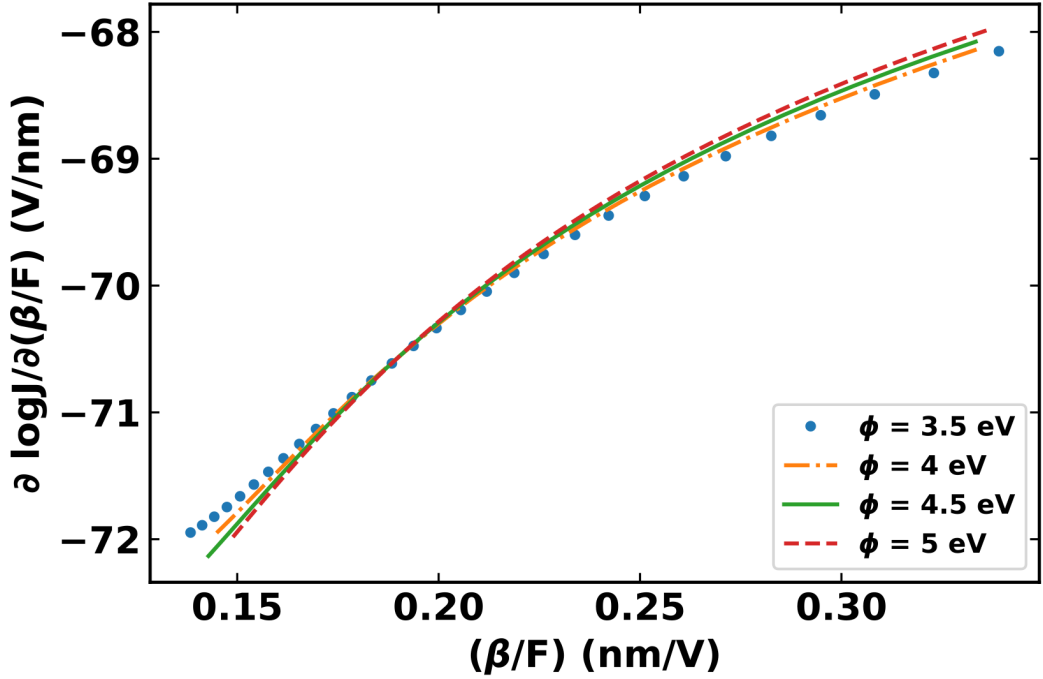}%
\caption{Slopes of the curves in figure \ref{figurereferee2}. $\beta$ = 0.6782 for $\phi$ = 3.5 eV. $\beta$ = 0.8339 for $\phi$ = 4 eV. $\beta$ = 1 for $\phi$ = 4.5 eV. $\beta$ = 1.176 for $\phi$ = 5 eV.\label{figurereferee3}}%
\end{figure}

In the introduction we stated that "the FN model needs 3 independent physical quantities, but the fit gives only 2 parameters". Actually, it can be shown\cite{charbonnier1962simple} that getting a good estimate of the area is possible even if $\phi$ and $\beta$ are unknown. So, in some sense field emission is more a problem with 2 physical quantities ($\phi$ and $\beta$) and a single parameter (the slope in the FN plot). One of the anonymous reviewer, suggested to add a non mandatory general comment to our article, that highlights this problem. So, we reproduce in this section his interesting point of view : 

There is a more general underlying reason why it is practically impossible for \textit{any} method (not only the Murphy-Good-plot one) to disentangle the work function from $\beta$. In Figure \ref{figurereferee1}, the current density - local field curves are calculated for various values of the work function W, following the "full Murphy-Good method", i.e., by numerically evaluating eq. \ref{eqfufllMG}. Figure \ref{figurereferee2} shows the same J curves plotted after rescaling the field and current density with an appropriate $\beta$ and S, chosen for each $\phi$ to fit to the curve with $\phi$=4.5 eV. It is obvious that the curves completely collapse to each other, yielding an RMS deviation of less than 0.5 $\%$ between the curves. In Figure \ref{figurereferee3}, the slope of these curves are plotted, again showing a deviation of less than 0.2$\%$ in the slopes.

This means that for an experimentalist to distinguish between these curves, i.e., disentangle the work function from
$\beta$, the following conditions should be fulfilled:
\begin{enumerate}
    \item The measured I-V curves or their derivatives should have a precision of the order 0.5$\%$
    \item The approximations of the Murphy-Good theory should yield an overall error of less than 0.5$\%$
    \item Approximating a constant effective emission area (constant ratio between total current and maximum current density) should be accurate within 0.5$\%$
\end{enumerate}
    
It is quite evident that the above requirements are practically implausible, at least with the state-of-heart
experimental methods. There might be theoretically a possibility to distinguish such curves by looking
into their second derivative (if that can be measured directly and precisely), however the reviewer still
opines that the aforementioned issues 2 and 3 (especially 3) are extremely hard to overcome.

\subsection{Putting an end to the tyranny of the straight line}

The remark of the reviewer in the previous section illustrates clearly the fact that very different field emitters can have very similar I-V curves. However in this final section, we would like to point out that changing the way of measuring field emission may have a positive impact.  Until now all the theoretical analysis were based on the idea of fitting a straight line and extracting a slope and an intercept. However, from an experimental point of view, this approach might not be so convenient, because of the presence of noise or drift in the current. So, we propose here a new approach requiring a single value of the applied voltage at the expense of the concomitant measurement of the current, its first and second voltage derivative\cite{ayari2021does}. If such experiment can be performed, it can be calculated from Eq. \ref{eqForb} that  $\kappa$ can be obtained by :  

\begin{equation}\label{maformule}
    \kappa = \frac{2IV\frac{dI}{dV}+IV^2\frac{d^2I}{dV^2}-(V\frac{dI}{dV})^2}{I^2}
\end{equation}

From such a measurement, if the simple model is correct, it is then possible to deduce the three other physical parameters ($\phi$, $S$ and $\beta$) by using Eq. \ref{giveK} for $\phi$ then Eq. \ref{theformule} for $\beta$ and Eq. \ref{eqForb} for S. The measurement can be performed by recording at the same time the DC current and AC current at the driving frequency of the lock-in (which gives the first derivative of the current) as well as at twice the driving frequency (which is proportional to the second derivative of the current). In practice this measurement is probably very challenging because it requires to subtract two very close numbers. Furthermore, although eq. \ref{maformule} is correct analytically, it is probably incorrect for real emitters.

\section{Conclusion}

We performed numerical calculations of the field emission current for several approximations of the Murphy and Good theory. We showed that the simple analytical formulae proposed in ref. \onlinecite{forbes2008call} gives significantly different results compared to the Murphy and Good analytical equation from which it was directly derived or compared to the full Murphy and Good theory at 0 K and 300 K. The predicted values of the work function by the simple model are not enough accurate for practical use and at this stage it is not clear if it could be improved. The main issue is that the central parameter $\kappa$ of the theory is not a simple parameter of the work function only. It depends also on the applied voltage and thus on $\beta$. The method proposed in ref. \onlinecite{forbes2008call} was a clever way to avoid the bending of the simple FN plot of the current calculated by the Murphy and Good model. Unfortunately, the Murphy and Good theory predicts also that the plot proposed in ref. \onlinecite{forbes2008call} is slightly bent and this bending cannot be neglected. It remains that if the Murphy and Good theory does not describe properly field emission, then $\kappa$ might still be an interesting empirical parameter.


%
%

%

\begin{acknowledgments}

The authors would like to thank both anonymous referees for their valuable comments.
\end{acknowledgments}

\section*{Data Availability}

The data that support the findings of this study are openly available in Zenodo at 

https://doi.org/10.5281/zenodo.5836944, Ref. \onlinecite{ayarizeno}.

\section*{Conflict of interest}

The authors have no conflicts to disclose.

\bibliography{refK}

\end{document}